\newcommand*{\ATLASLATEXPATH}{latex/}
\documentclass[cernpreprint, texlive=2011, txfonts, UKenglish]{\ATLASLATEXPATH atlasdoc}

\usepackage[subfigure]{\ATLASLATEXPATH atlaspackage}
\usepackage{\ATLASLATEXPATH atlasbiblatex}

\usepackage{multirow}

\usepackage{\ATLASLATEXPATH atlasphysics}

\graphicspath{{logos/}{figures/}}

\newcommand{\intlumi}{3.2~fb${}^{-1}$}
\newcommand{\Pythiasix}{{\textsc{Pythia6}}\xspace}
\newcommand{\Pythiaeight}{{\textsc{Pythia8}}\xspace}
\newcommand{\Sherpa}{{\textsc{Sherpa~2.1}}\xspace}
\newcommand{\SherpaNoVer}{{\textsc{Sherpa}}\xspace}
\newcommand{\Powheg}{{\textsc{POWHEG}}\xspace}
\newcommand{\charybdis}{{\textsc{Charybdis2 1.0.4}}\xspace}
\newcommand{\charybdisNoVer}{{\textsc{Charybdis2}}\xspace}
\newcommand{\geant}{{\textsc{Geant4}}\xspace}
\newcommand{\sumpt}{\ensuremath{\sum \pt}\xspace}
\newcommand{\MD}{\ensuremath{M_{\mathrm{D}}}\xspace}
\newcommand{\MTH}{\ensuremath{M_{\mathrm{th}}}\xspace}

\newcommand{\Wjets}{\ensuremath{\Wboson{+}\text{jets}}}
\newcommand{\Zjets}{\ensuremath{\Zboson{+}\text{jets}}}

\newcommand{\SROne}{{\texttt{SR-2TeV}}}
\newcommand{\SRTwo}{{\texttt{SR-3TeV}}}

\newcommand{\figdisclaimer}{{The \ttbar, \Wjets\ and \Zjets\ backgrounds are normalised by the factors \scaleFactorTTBARVal, \scaleFactorWJETSVal\ and \scaleFactorZJETSVal\ as obtained from the background likelihood fit. The single-top-quark and diboson background normalisations are taken from the simulation.  The multijet background is obtained using a data-driven method. Additionally, the likelihood fit may constrain nuisance parameters for certain systematic uncertainties, altering the normalisation and shape of some of the distributions. }}

\newcommand{\figexplanationCore}[1]{{The data are shown as points with error bars; all expected backgrounds are shown as stacked coloured histograms, with the total background uncertainty shown as a shaded band. #1The lower panels show the ratio of the data to the expected background.}}
\newcommand{\figexplanation}{{\figexplanationCore{}}}
\newcommand{\figexplanationwithsignal}{{\figexplanationCore{Two representative signal distributions for rotating black holes with $n =6$ are overlaid to illustrate the signal properties. }}}

\newcommand{\tableerrorsstatement}{The errors shown are the statistical plus systematic uncertainties.
The uncertainty in the total background count includes correlations between nuisance parameters and so does not reflect a quadrature sum of the uncertainties in the individual background components.}

\newcommand{\expectedSrHighCo}{{$ 9.9 \pm 1.4$}}
\newcommand{\expectedSrLowwCo}{{$ 181 \pm 11$}}

\newcommand{\observedSrHighCo}{{13}}
\newcommand{\observedSrLowwCo}{{192}}

\newcommand{\modelIndependentBoundNinetyFiveUnits}{{fb}}

\newcommand{\modelIndependentBoundNinetyFiveLowwCo}{{12.1}}

\newcommand{\modelIndependentBoundNinetyFiveHighCo}{{3.4}}

\newcommand{\scaleFactorTTBARVal}{{0.95}}
\newcommand{\scaleFactorTTBARShortErr}{{\pm0.08}} 
\newcommand{\scaleFactorWJETSVal}{{0.81}}
\newcommand{\scaleFactorWJETSShortErr}{{\pm0.07}} 
\newcommand{\scaleFactorZJETSVal}{{1.01}}
\newcommand{\scaleFactorZJETSShortErr}{{\pm0.08}} 

\hypersetup{pdftitle={ATLAS paper},pdfauthor={The ATLAS Collaboration}}

\AtlasTitle{
Search for TeV-scale gravity signatures in high-mass final states with leptons and jets with the ATLAS detector at $\sqrt{s}=13$ TeV
}

\author{The ATLAS Collaboration}

\AtlasRefCode{EXOT-2015-14}

\PreprintIdNumber{CERN-EP-2016-109}

\AtlasJournal{Phys.\ Lett.\ B}

\AtlasAbstract{%
  A search for physics beyond the Standard Model, in final states with at least one high transverse momentum charged lepton (electron or muon) and two additional high transverse momentum leptons or jets, is performed using \intlumi\ of proton--proton collision data recorded by the ATLAS detector at the Large Hadron Collider in 2015 at $\sqrt{s} = 13$~TeV.
  The upper end of the distribution of the scalar sum of the transverse momenta of leptons and jets is sensitive to the production of high-mass objects.
  No excess of events beyond Standard Model predictions is observed.
  Exclusion limits are set for models of microscopic black holes with two to six extra dimensions.
}

\AtlasCoverSupportingNote{ATL-COM-PHYS-2016-247}{https://cds.cern.ch/record/2139302}
\AtlasCoverAnalysisTeam{
Benjamin~Hylton~Brunt~(Cambridge),
Teng-Jian~Khoo~(Geneva),
Christopher~Gorham~Lester~(Cambridge),
Michael~Andrew~Parker~(Cambridge),
Sarah~Louise~Williams~(NIKHEF)
}

\AtlasCoverEdBoardMember{Jean-Francois~Grivaz~(Orsay~LAL)~(chair)}
 \AtlasCoverEdBoardMember{Domizia~Orestano~(Roma Tre)}
 \AtlasCoverEdBoardMember{Antonio~Boveia~(CERN)}

\AtlasCoverEgroupEditors{atlas-EXOT-2015-14-editors@cern.ch}

\AtlasCoverEgroupEdBoard{atlas-EXOT-2015-14-editorial-board@cern.ch}

\begin{document}

\maketitle


\section{Introduction}
\label{sec:intro}

Models of TeV-scale gravity postulate that the fundamental scale of gravity, \MD, in a higher-dimensional space-time is much lower than is measured in our four-dimensional space-time. In large extra-dimensional models (e.g.~the model proposed by Arkani-Hamed, Dimopoulos and Dvali (ADD)
\cite{ArkaniHamed:1998rs,Antoniadis:1998ig}) there are $n$ additional flat extra dimensions, assumed to be compactified on a torus with a common radius much larger than $1/\MD$. Another class of models (e.g.\ that of Randall and Sundrum (RS) \cite{Randall:1999ee, Randall:1999vf}) uses one extra dimension in a highly warped anti-de-Sitter space. Both of these types of model can address the large difference between the scale of electroweak interactions, ${\cal O}(0.1$~TeV), and that of gravity, the Planck scale, $ M_\mathrm{Pl}={\cal O}(10^{16}$~TeV), in a natural way. Interesting signatures are expected in these models in the form of non-perturbative gravitational states such as microscopic black holes \cite{DimopoulosLandsbergBHatLHC,Giddings:2001bu}. Such final states could be produced in proton--proton ($pp$) interactions at the Large Hadron Collider (LHC) \cite{LHCProposalPaper}. In the absence of a full theory of quantum gravity, predictions for production cross-sections and decays of black holes rely on semi-classical approximations which are expected to be valid if the mass of the black hole is well above \MD and also higher than the Hawking temperature~\cite{Hawking:1974sw}.
A strong rise in the production rate of such states is expected when the energy scale of the interactions reaches the order of \MD . Since the gravitational interaction couples to the energy-momentum tensor rather than gauge quantum numbers, final states are expected to be populated ``democratically'', according to the number of available Standard Model degrees of freedom. For this reason, it is expected that a significant fraction of final states would contain leptons. This search exploits this feature to enhance the signal contribution in comparison with the dominant background at the LHC, which arises from quark and gluon scattering processes forming hadronic final states.
Final states with at least three high transverse momentum (\pt) objects are selected, of which at least one must be an electron or muon (leptons in what follows) and the others can be either leptons or hadronic jets. The discriminating variable used in this search, \sumpt, is the scalar sum of the transverse momenta of high \pt\ objects in an event. The signal is expected to appear at high \sumpt.
Searches by ATLAS~\cite{Aad:2011bw, Aad:2012ic, Aad:2013bw, Aad:2014gka} and CMS~\cite{Khachatryan:2010wx,CMSBlackHoles12fb} during Run 1 of the LHC did not reveal any significant excesses over expected background levels.
An ATLAS analysis~\cite{Aad:2015mzg} of Run-2 data at 13~TeV also found no evidence of new effects in multijet final states.
This work extends the reach of the analysis in Ref.~\cite{Aad:2014gka}, performed at a centre-of-mass energy of 8~TeV, with \intlumi\ of data recorded by ATLAS in 2015 at 13~TeV.
This search is potentially sensitive to other forms of new physics at high-mass and involving the electroweak sector.

\section{ATLAS detector}
\label{sec:detector}

ATLAS~\cite{atlas-detector} is a multipurpose detector with a forward-backward symmetric cylindrical
geometry and nearly 4$\pi$ coverage in solid angle.\footnote{ATLAS uses a right-handed coordinate system with its origin at the nominal interaction point (IP) in the centre of the detector and the $z$-axis along the beam direction. The $x$-axis points from the IP to the centre of the LHC ring, and the $y$-axis points upward. Cylindrical coordinates $(r,\phi)$ are used in the transverse plane, $\phi$ being the azimuthal angle around the beam direction. The pseudorapidity is defined in terms of the polar angle $\theta$ as $\eta=-\ln\tan(\theta/2)$. Object separations are measured using $\Delta R=\sqrt{(\Delta\phi)^2+(\Delta\eta)^2}$.} The inner detector (ID) utilises fine-granularity pixel and microstrip detectors over the pseudorapidity range $|\eta|< 2.5$ to provide precise track parameter and secondary vertex measurements. For {Run 2} of the LHC, a new pixel layer has been added at a radius of 3.3~cm \cite{Capeans:1291633}. A gas-filled straw-tube tracker complements the silicon tracker at larger radii.  The tracking detectors are immersed in a $2$~T magnetic field produced by a thin superconducting solenoid.  The electromagnetic (EM) calorimeters employ lead absorbers and use liquid argon as the active medium.  The barrel EM calorimeter covers $|\eta|<1.5$ and the end-cap EM calorimeters cover $1.4 < |\eta| <3.2$.  Hadronic calorimetry in the region $|\eta|< 1.7$ is performed using steel absorbers with scintillator tiles as the active medium.  Liquid-argon calorimetry with copper absorbers is used in the hadronic end-cap calorimeters, which cover the region $1.5 < |\eta| < 3.2$. The forward calorimeters ($3.1 < |\eta| < 4.9$) use copper and tungsten as absorber with liquid argon as active material. The muon spectrometer (MS) measures the deflection of muon trajectories in the region $|\eta|< 2.7$, using three stations of precision drift tubes (with cathode strip chambers in the innermost station for $|\eta| > 2.0$)  located in a toroidal magnetic field of approximately $0.5$~T and $1$~T in the central and end-cap regions, respectively. The muon spectrometer is also instrumented with separate trigger chambers covering $|\eta|< 2.4$.  Events are selected using a first-level trigger implemented in custom electronics, which reduces the event rate with a design value of $100$~kHz using a subset of detector information~\cite{ATL-DAQ-PUB-2016-001}. Software algorithms with access to the full detector information are then used to yield a recorded event rate of about $1$~kHz.

\section{Analysis}
\subsection{Signal simulation}
Signal samples are generated by using the \charybdis generator \cite{Frost:2009cf} to simulate the production and decay of rotating black holes in models with $n=2$, 4 and 6 extra dimensions and values of \MD ranging from 2~TeV to 5~TeV.
Black holes are assumed to be produced over a continuous range of mass values above a threshold \MTH, set so as to avoid the theoretical uncertainties associated with the region close to \MD.
The analysis is guided by two benchmark signal models, the first of which has $\MD = 2~\text{TeV}$ and $\MTH = 7~\text{TeV}$, resulting in a cross-section of $0.72~\text{pb}$.
The second has $\MD = 4~\text{TeV}$, $\MTH = 6~\text{TeV}$, and a cross-section of $0.93~\text{pb}$.
In these simulations, no initial-state gravitational radiation is permitted, while the final decay of the black-hole remnant produces a variable number of particles, whose multiplicity is drawn from a Poisson distribution in accordance with the \charybdisNoVer\ default.
The CTEQ6L1 parton distribution functions (PDFs) used are taken from Ref.~\cite{Pumplin:2002vw}, while the final-state fragmentation and parton showering is modelled using \Pythiaeight~\cite{pythia8}.
The detector response is modelled using a fast simulation of the response of the calorimeters~\cite{Aad:2010ah} and \geant~\cite{geant4} for other parts of the detector.
Events from minimum-bias interactions are also simulated with \Pythiaeight . They are overlaid on the simulated signal and background events according to the luminosity profile of the recorded data. Interactions within the same bunch crossing as the hard-scattering process and in neighbouring bunch crossings are both simulated and are referred to as pile-up.

\subsection{Event selection}
Events are selected from a sample with an integrated luminosity of $3.2 \pm 0.2$~fb$^{-1}$.
The luminosity estimate is derived following the same methodology as that detailed in Ref.~\cite{Aad:2013ucp}, from a calibration of the luminosity scale using a pair of $x$--$y$
beam-separation scans performed in August 2015.
The event selection uses the lowest-threshold single-lepton triggers available in each data-taking period with good operational conditions.
The single-electron trigger uses a minimum threshold of $\ET=60~\text{GeV}$.
The minimum threshold used for the single-muon trigger is $\pt=50~\text{GeV}$.
All the final-state objects are required to satisfy basic criteria to ensure that they are well reconstructed and originate from the primary interaction.
Candidate electrons and muons are required to have $p_\mathrm{T}>10$~GeV and pseudorapidity $|\eta|<2.47$ (electrons) or $|\eta|<2.5$ (muons).
They are also required to satisfy baseline identification criteria (the ``Loose'' operating point of Ref.~\cite{ATL-PHYS-PUB-2015-041} for electrons and the ``Medium'' criteria of Ref.~\cite{Aad:2016jkr} for muons).
 Jets of hadrons are reconstructed using the anti-$k_t$ algorithm with a radius parameter of 0.4 \cite{Cacciari:2008gp} and are required to be of at least ``loose'' quality \cite{ATLAS-CONF-2015-029} and to have a calibrated \cite{ATL-PHYS-PUB-2015-015} $p_\mathrm{T}>20$~GeV and $|\eta|<2.8$.
 Jets containing $b$-hadrons are identified using the ``$b$-tagging'' techniques described in Refs. \cite{ATL-PHYS-PUB-2015-022,ATL-PHYS-PUB-2015-039}.
To avoid double-counting of reconstructed objects, electrons sharing an inner detector track with a muon are removed. Following this, jet candidates that are not $b$-tagged are removed if they are within $\Delta R<0.2$ of an electron candidate. 
Finally, any lepton candidate within $\Delta R < 0.4$ of a surviving jet candidate that is not tagged as originating from pile-up~\cite{Aad:2015ina} is removed.
The remaining electrons are required to satisfy the ``Tight'' operating point of Ref.~\cite{ATL-PHYS-PUB-2015-041}.
Leptons are required to be isolated from other activity using a relatively loose criterion designed to pass 99\% of leptons from $Z$ decays~\cite{Aad:2016jkr,ATLAS-CONF-2014-032}.
 Events are sorted into electron and muon channels according to the flavour of the highest $p_\mathrm{T}$ lepton. Two signal regions (SRs) are defined, requiring a leading lepton with $p_\mathrm{T}>100$~GeV and at least two other objects (leptons or jets) with $p_\mathrm{T}>100$~GeV, with $\sumpt>2$~TeV or 3~TeV, where \sumpt includes all objects in the event with $p_\mathrm{T}>60$~GeV. The first signal region (named \SROne) allows the search to cover the parameter space near the existing limits, while the second (named \SRTwo) provides sensitivity at the highest \sumpt accessible.
The \SRTwo\ selection gives $\text{efficiency} \times \text{acceptance}$ values for the benchmark signal models of 19\% (for the model at $\MD = 2~\text{TeV}$, $\MTH = 7~\text{TeV}$) and
8\% (for the model at $\MD = 4~\text{TeV}$, $\MTH = 6~\text{TeV}$).

\subsection{Backgrounds}
The dominant backgrounds originate from $W$ and $Z$ boson production associated with hadronic jets (\Wjets\ and \Zjets) and from \ttbar\ production.
For these backgrounds, the distributions in kinematic quantities are predicted by Monte Carlo (MC) simulations, which are normalised to data in dedicated control regions (CRs).
Each CR uses selections which enhance the contribution of the relevant background while maintaining a negligible expected signal contribution.
Single-top-quark and diboson production processes give small contributions that are estimated directly from simulations, with normalisations taken from Refs.~\cite{hathor,hathor2}
and from the generator, respectively. The bosonic background processes are simulated using \Sherpa~\cite{Gleisberg:2008ta}, while \Powheg~\cite{Nason:2004rx,powhegPaper,Alioli:2010xd} in conjunction with \Pythiasix~\cite{Sjostrand:2006za} is used for top quark production processes. All these background simulations use the CT10 PDF set~\cite{Lai:2010vv}.  The detector response is modelled using \geant. The electron channel also contains background events from hadronic jets which are incorrectly reconstructed as electrons. This background, called ``multijet'', is estimated from the data using a sample of events selected with loosened identification criteria using the Matrix Method~\cite{Aad:2010ey}. The rate of background muons from hadronic jets is negligible.

{The {background} CR selection criteria are summarised in Table~\ref{tab:controlRegionSummary}.}
All of the CRs select events with $750<\sumpt<1500$~GeV, including at least three objects with $p_\mathrm{T}>60$~GeV of which one is required to be a lepton. The $Z$+jets CR additionally requires exactly two leptons with the same flavour and opposite charge and an invariant mass $m_{\ell\ell}$ in the range 80--100~GeV. The $W$+jets CR requires events with exactly one lepton {and a missing transverse momentum} $E_\mathrm{T}^\mathrm{miss}$
\cite{ATL-PHYS-PUB-2015-023}
 exceeding $60$~GeV. In this CR, in order to suppress background from top quark production, none of the jets may be $b$-tagged.
 The \ttbar\ CR also requires exactly one lepton, but there must be at least four jets of which at least two are $b$-tagged. In order to use information about the shape of the $\sumpt$ distribution to more accurately constrain the normalisation of the $W$+jets, $Z$+jets and $t\bar{t}$ backgrounds in the SRs, each control region is divided into three 250-GeV-wide bins.

\begin{table}[tb]
	\begin{center}
		\begin{tabular}{|c|c|c|c|c|}
			\hline
			\multirow{2}{*}{\textbf{Selection}} & \multicolumn{3}{c|}{\textbf{Control Regions}} & \multirow{2}{*}{\textbf{Signal regions}} \\ \cline{2-4}
      & \textbf{\Zjets} & \textbf{\Wjets} & \rule{0pt}{2ex} \textbf{\ttbar} &  \\ \hline

			$\sumpt$ &\multicolumn{3}{c|}{$750$--$1500$~GeV} & $>2000(3000)$~GeV  \\ \hline
      Number of objects  & \multicolumn{3}{c|}{$\geq$ 3 objects}  & $\geq$ 3 objects \\
      (leptons or jets)  & \multicolumn{3}{c|}{with $\pt > 60$ GeV} & with $\pt > 100$ GeV  \\
      \hline
      Leading lepton     & \multicolumn{3}{c|}{Isolated } & Isolated \\
      (electron or muon) & \multicolumn{3}{c|}{with $\pt > 60$ GeV}     & with $\pt > 100$ GeV  \\
       \hline
      $m_{\ell\ell}$           &  80--100 GeV     & \multicolumn{2}{c|}{n/a} & \multirow{2}{*}{n/a} \\ \cline{0-3}
      \rule{0pt}{4mm} $\displaystyle \met$ &  n/a               & $> 60$~\GeV & n/a        &                      \\ \cline{0-4}
			\multirow{2}{*}{Number of leptons} &  $= 2$, opposite sign & \multicolumn{2}{c|}{\multirow{2}{*}{$= 1$}} & \multirow{2}{*}{$\geq 1$} \\
			                                   &  same flavour             & \multicolumn{1}{c}{} &                          &  \\ \cline{0-4}
      Number of $b$-tagged jets &  n/a & $= 0$ & $\geq 2$ & \multirow{2}{*}{n/a} \\ \cline{0-3}
			Number of jets &  \multicolumn{2}{c|}{n/a} & $\geq 4$ & \\
      \hline

		\end{tabular}
	\end{center}
	\caption{Definitions of the signal regions and of the control regions used in the estimate of the \Wjets, \Zjets\ and~\ttbar~backgrounds.}
	\label{tab:controlRegionSummary}
\end{table}

\subsection{Systematic uncertainties}
The systematic uncertainties in the signal and backgrounds include those due to the limited numbers of simulated events
and to the measurement of integrated luminosity. Experimental uncertainties arising from the trigger efficiencies, lepton identification and reconstruction procedures, the $b$-tagging algorithm and the energy calibration of leptons and jets, as well as effects from the jet energy resolution, are also taken into account.
Potential mismodelling by the MC simulations of the $W$+jets, $Z$+jets and \ttbar\ backgrounds is quantified by comparing the nominal against alternative simulated samples and PDF sets.  For the $W$+jets and $Z$+jets backgrounds, simulated by \SherpaNoVer, the default renormalisation, factorisation and resummation scales are doubled or halved. The matrix element and parton shower are matched using the CKKW~\cite{Catani:2001cc} scheme, for which the default scale of 20~GeV is changed to 15~GeV and to 30~GeV.  For \ttbar, uncertainties in the hard scatter and fragmentation are estimated by comparison with alternative generators and parton shower models. Variations of the renormalisation scale and of the amount of initial- and final-state radiation are performed within the nominal generator.  Since the overall normalisations of the backgrounds are well constrained by the fits to the data described below, only variations in shape as a function of $\sumpt$ are relevant.
The systematic uncertainty in the predicted yields in both channels of  \SROne\ and \SRTwo\ is dominated by the limited sizes of the Monte Carlo samples.
The total uncertainties in the SRs are mainly of statistical origin.

\FloatBarrier
\section{Results}
\label{sec:result}
\begin{figure*}[ptb]
  \begin{center}
    \subfigure[]{
      \includegraphics[width=0.42\textwidth]{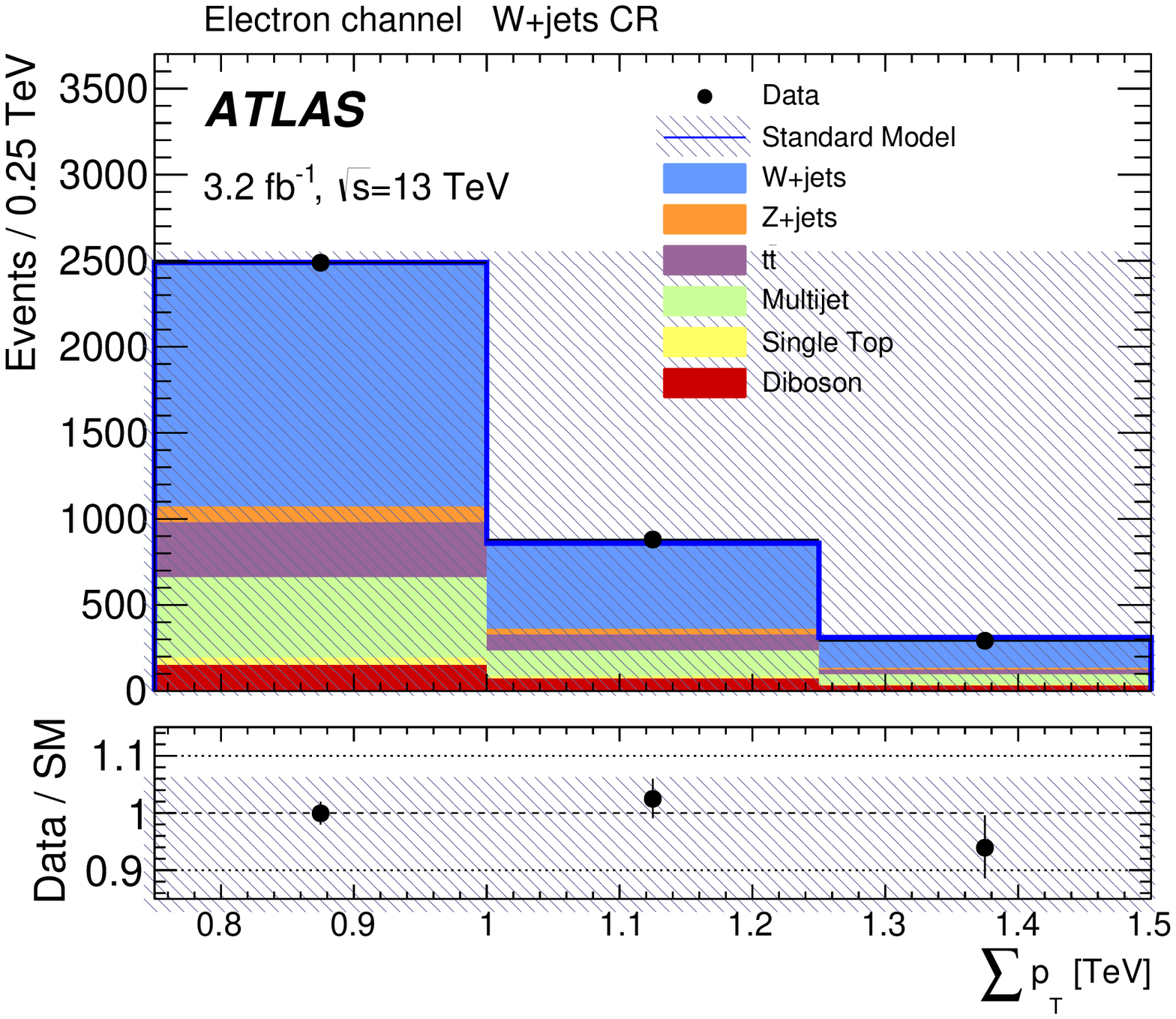}
      \label{fig:SumpTWCR:elChan}
    }
    \subfigure[]{
      \includegraphics[width=0.42\textwidth]{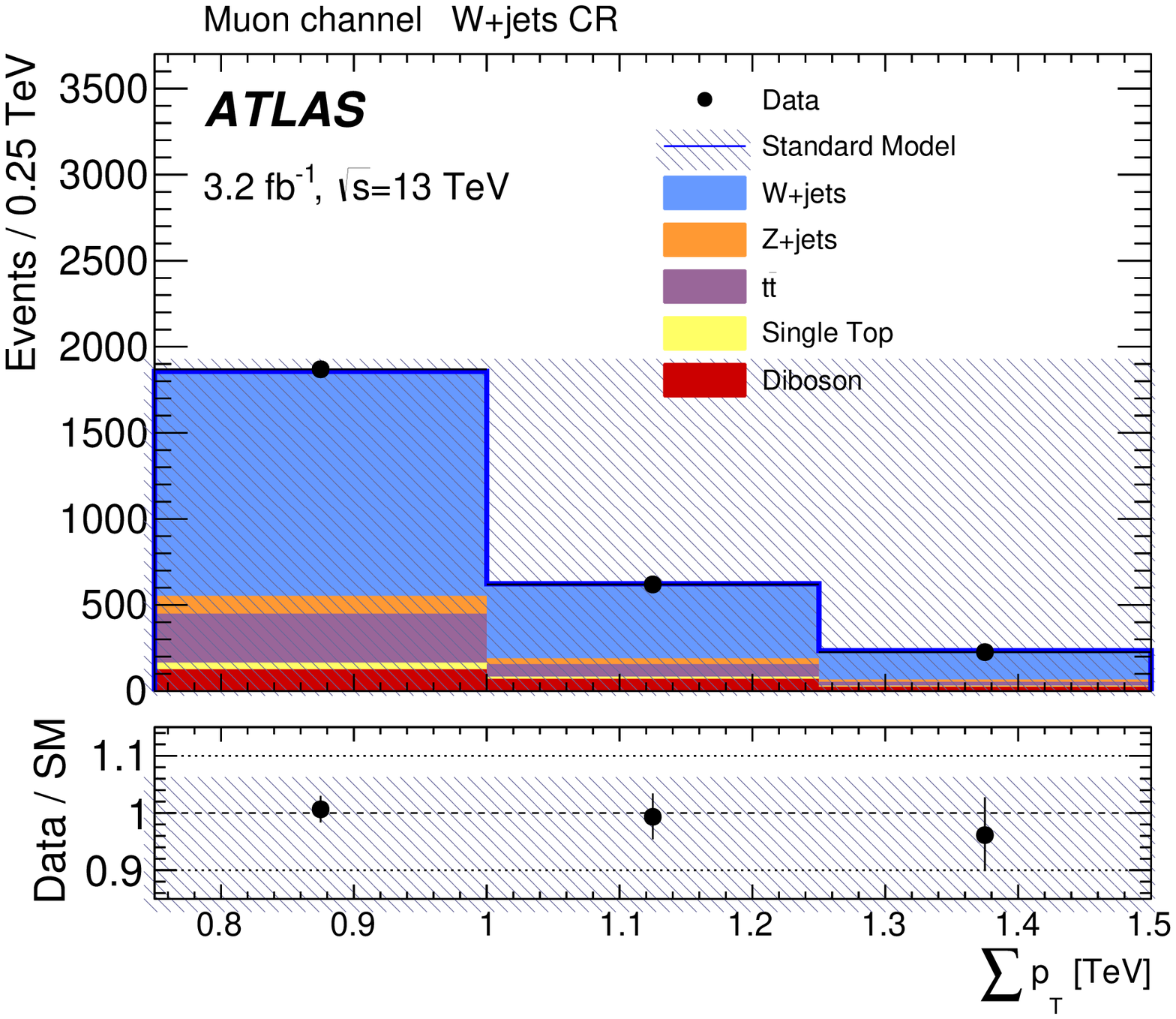}
      \label{fig:SumpTWCR:muChan}
    }
    \subfigure[]{
      \includegraphics[width=0.42\textwidth]{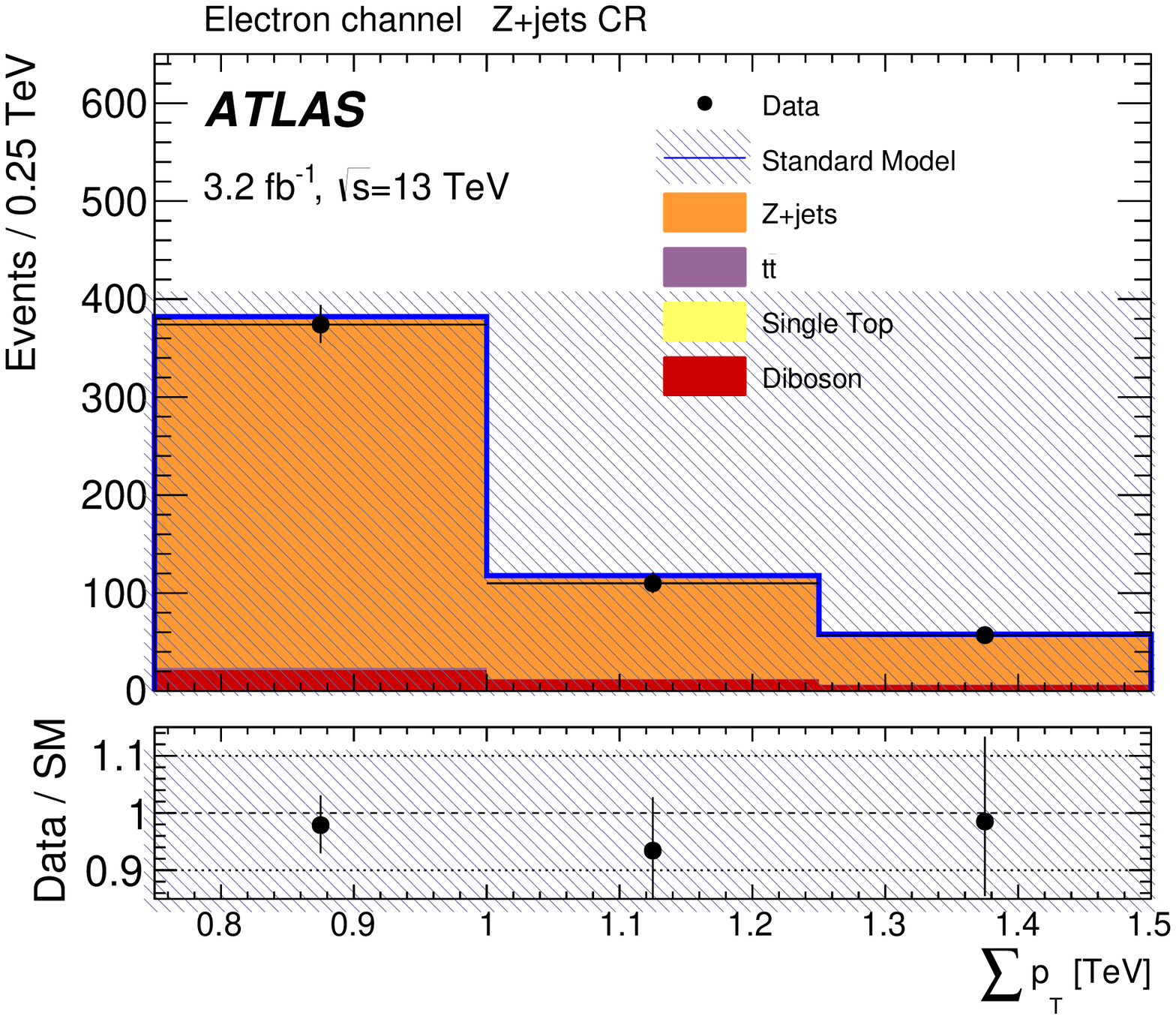}
      \label{fig:SumpTZCR:elChan}
    }
    \subfigure[]{
      \includegraphics[width=0.42\textwidth]{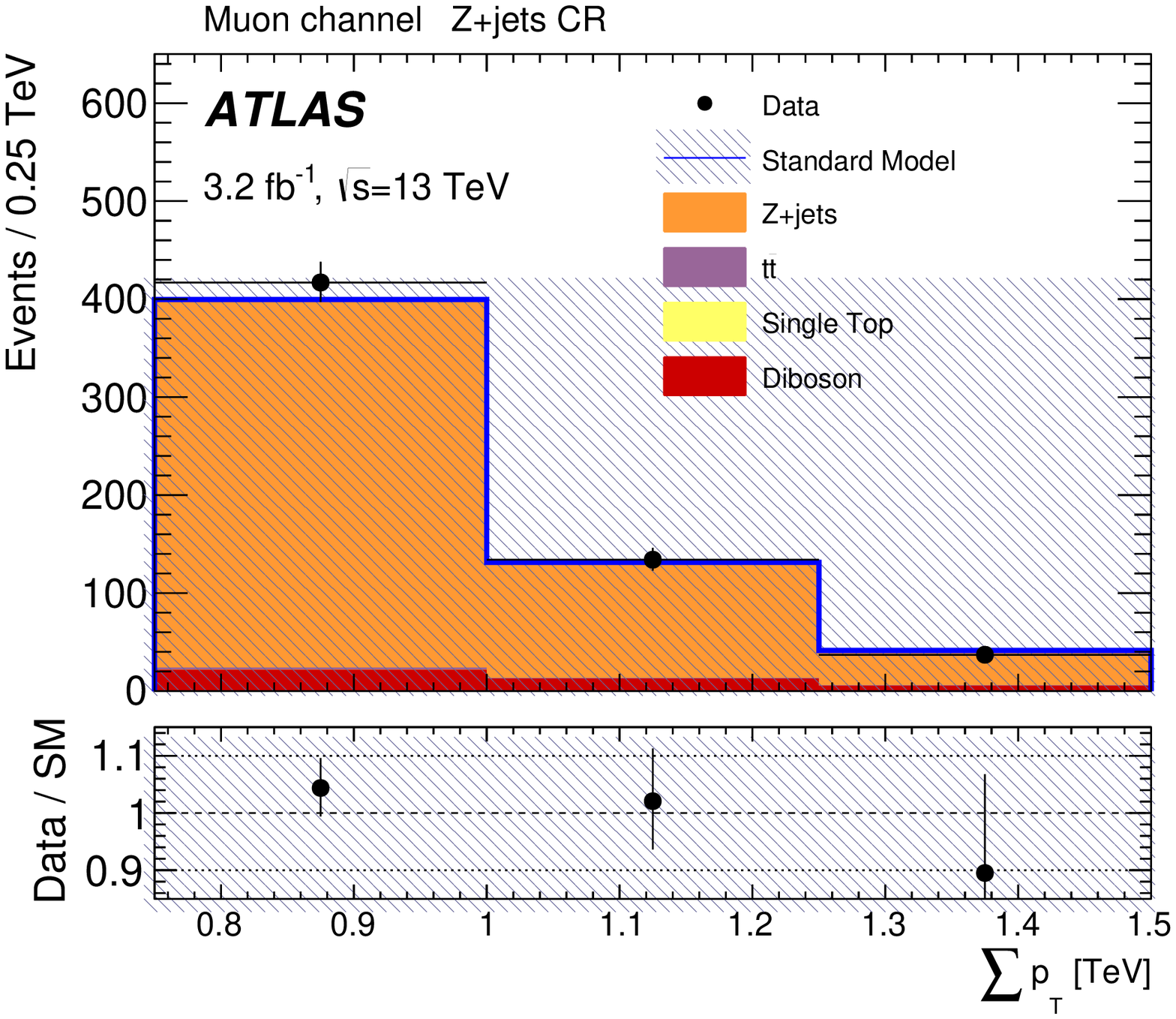}
      \label{fig:SumpTZCR:muChan}
    }
    \subfigure[]{
      \includegraphics[width=0.42\textwidth]{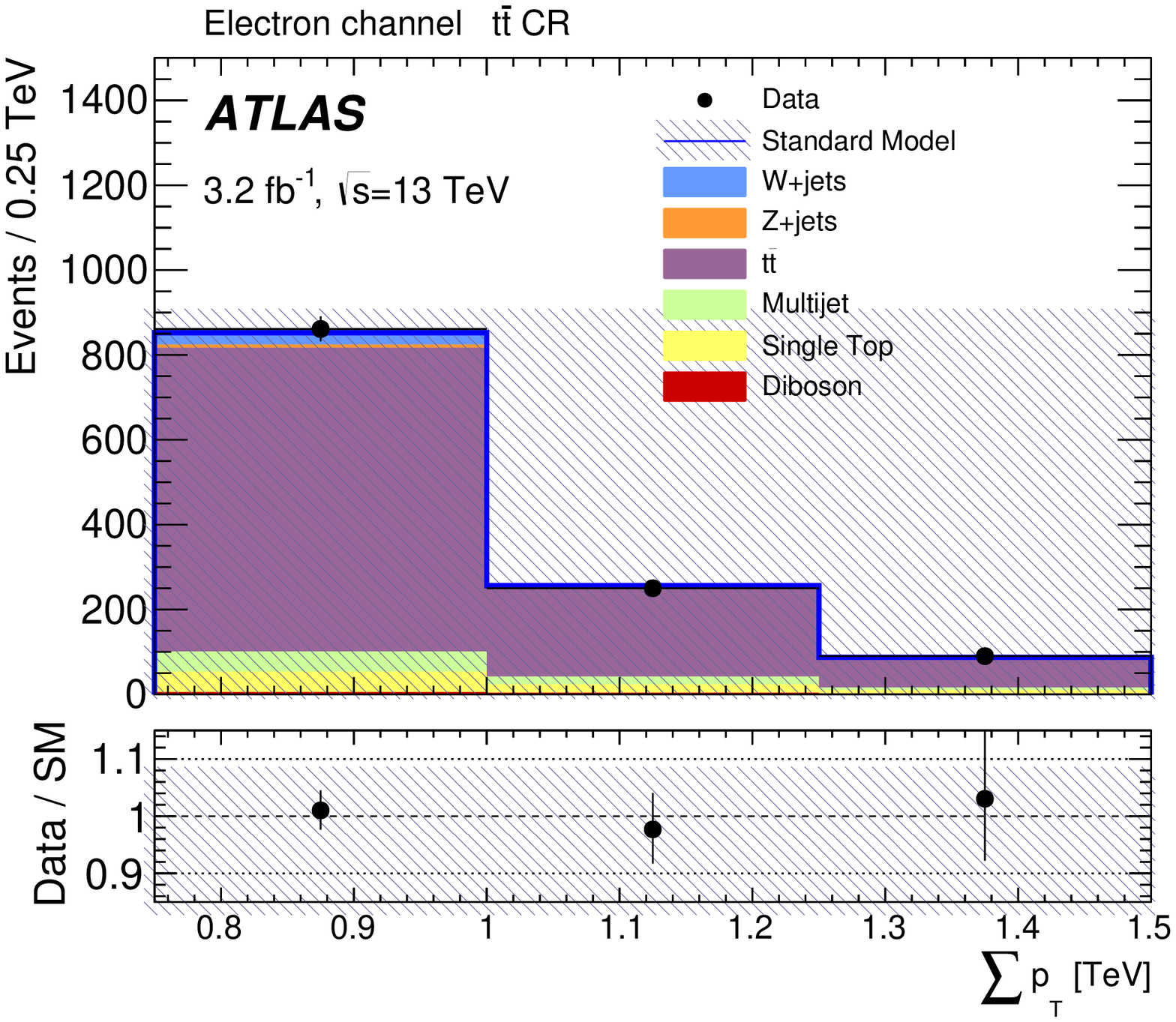}
      \label{fig:SumpTTopCR:elChan}
    }
    \subfigure[]{
      \includegraphics[width=0.42\textwidth]{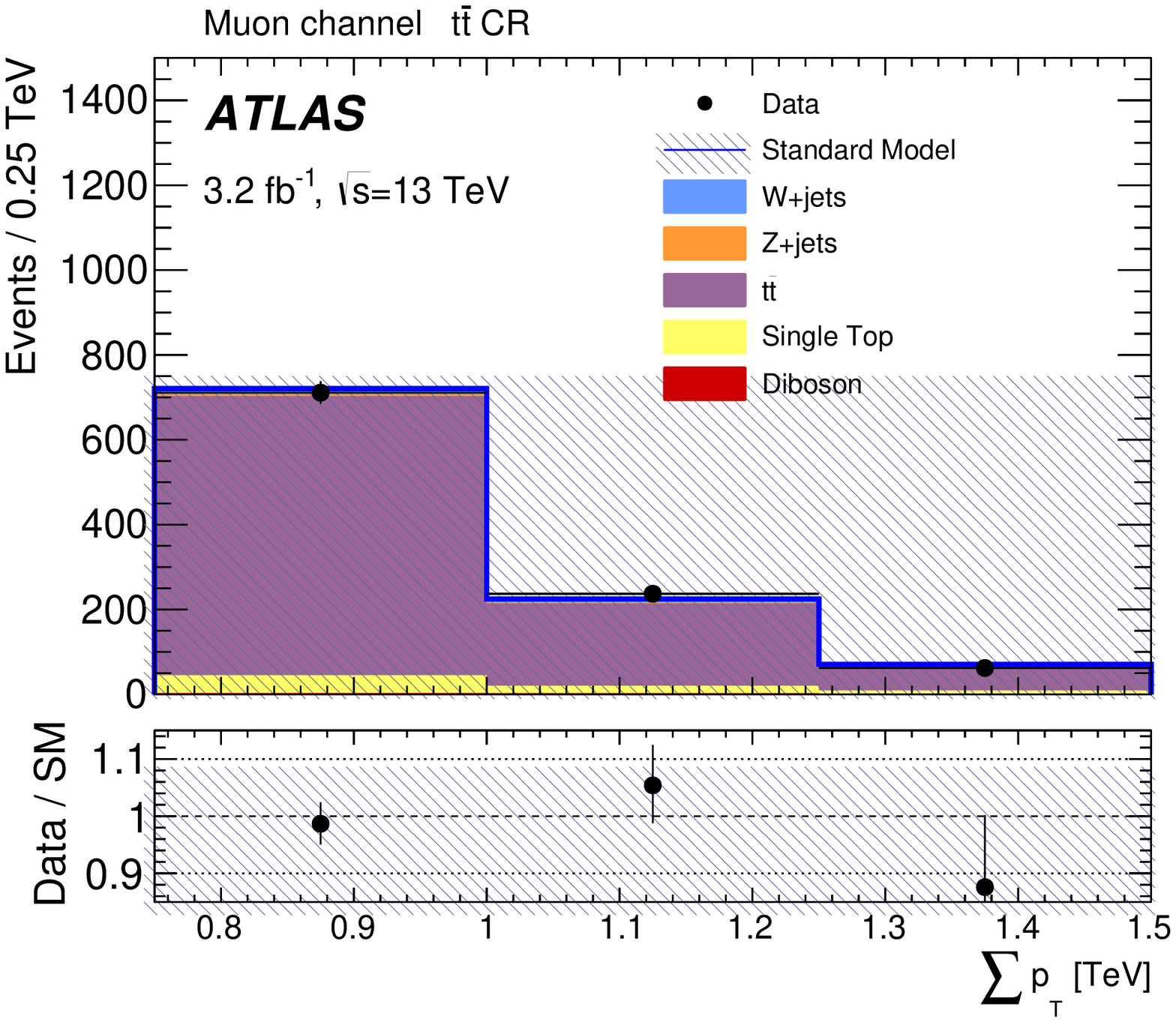}
      \label{fig:SumpTTopCR:muChan}
    }
    \caption{The \sumpt distribution in each of the control regions.
      The \Wjets\ CR is shown in \subref{fig:SumpTWCR:elChan} and \subref{fig:SumpTWCR:muChan}, the \Zjets\ CR in \subref{fig:SumpTZCR:elChan} and \subref{fig:SumpTZCR:muChan}, and the \ttbar\ CR in \subref{fig:SumpTTopCR:elChan} and \subref{fig:SumpTTopCR:muChan}.
      The electron channel is shown in \subref{fig:SumpTWCR:elChan}, \subref{fig:SumpTZCR:elChan} and \subref{fig:SumpTTopCR:elChan},
      and the muon channel in \subref{fig:SumpTWCR:muChan}, \subref{fig:SumpTZCR:muChan} and \subref{fig:SumpTTopCR:muChan}.
    \figexplanation\ \figdisclaimer}
    \label{fig:SumpTCRs}
  \end{center}
\end{figure*}

 Results are extracted from profile likelihood fits using three background normalisation parameters for the $W+$jets, $Z+$jets and $t\bar{t}$ backgrounds.
 These normalisation parameters are freely floating in the fits.
 Nuisance parameters are included in the fits to describe the systematic uncertainties, taking into account the correlations across the processes and regions involved in each fit.
A background likelihood fit to all control regions of both lepton channels, assuming no signal contribution, is used to predict the expected yields in validation regions (VRs) and to test the hypothesis that the data is well described with no signal in these regions.
The VRs are defined using the same event selections as the signal regions, but in the range $1500<\sumpt<2000$~GeV.
As in the CRs, any signal contamination in the VRs is expected to be small, based on previous analyses \cite{Aad:2014gka} and on signal simulations.
Comparisons between the data and the predictions in the control regions, where the background predictions are adjusted by the background likelihood fit, may be seen in Figure~\ref{fig:SumpTCRs}.
The MC simulation provides a good description of the CR data, with scale factors of
$\scaleFactorWJETSVal \scaleFactorWJETSShortErr$,
$\scaleFactorZJETSVal \scaleFactorZJETSShortErr$ and
$\scaleFactorTTBARVal \scaleFactorTTBARShortErr$
for $W$+jets, $Z$+jets and $t\bar{t}$ respectively.
No significant deviation from the background prediction is observed in the VRs.

\begin{figure*}[tbp]
  \begin{center}
    \subfigure[]{
      \includegraphics[width=0.45\textwidth]{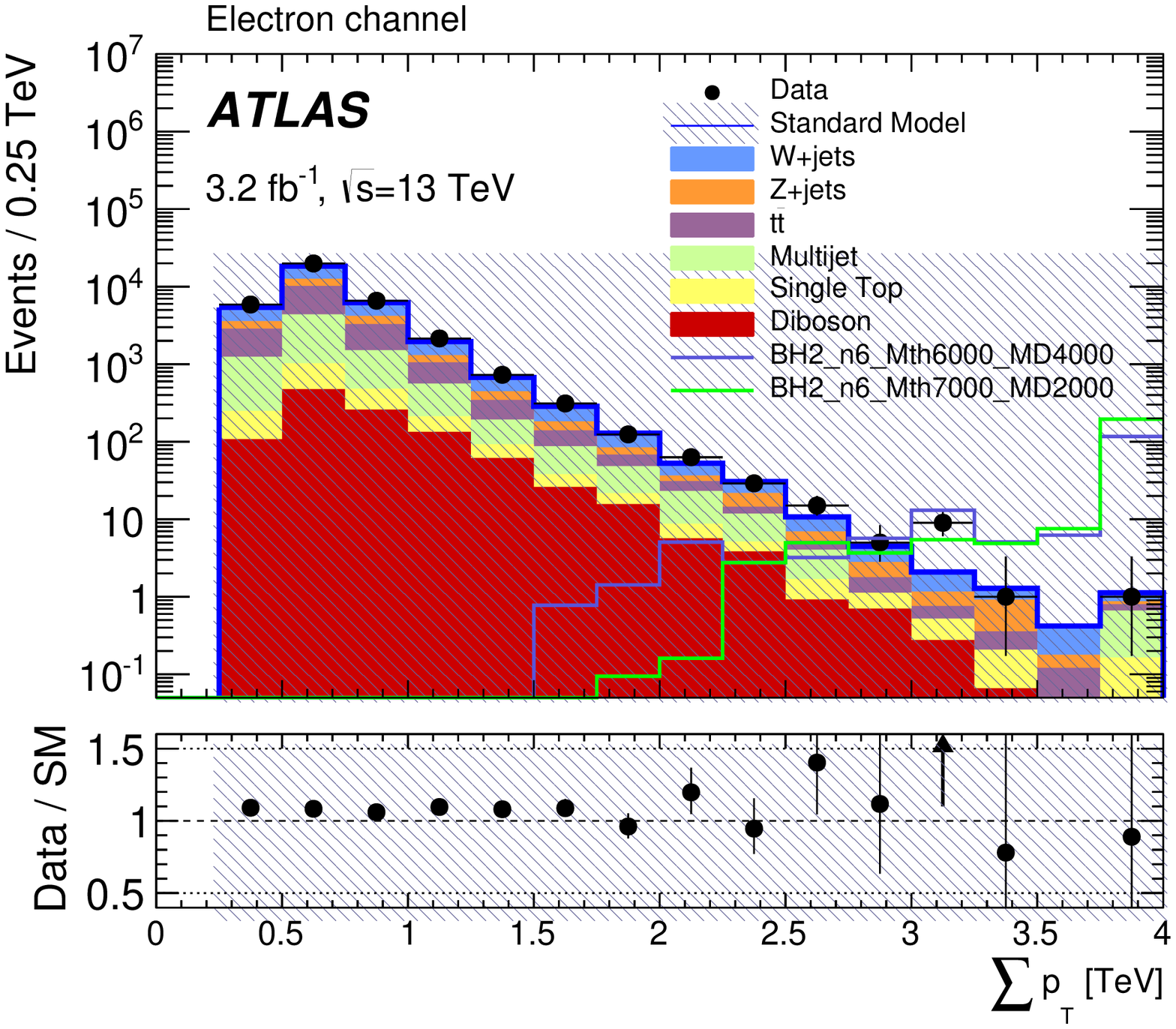}
      \label{fig:finalsumpt:elChan}
    }
    \subfigure[]{
      \includegraphics[width=0.45\textwidth]{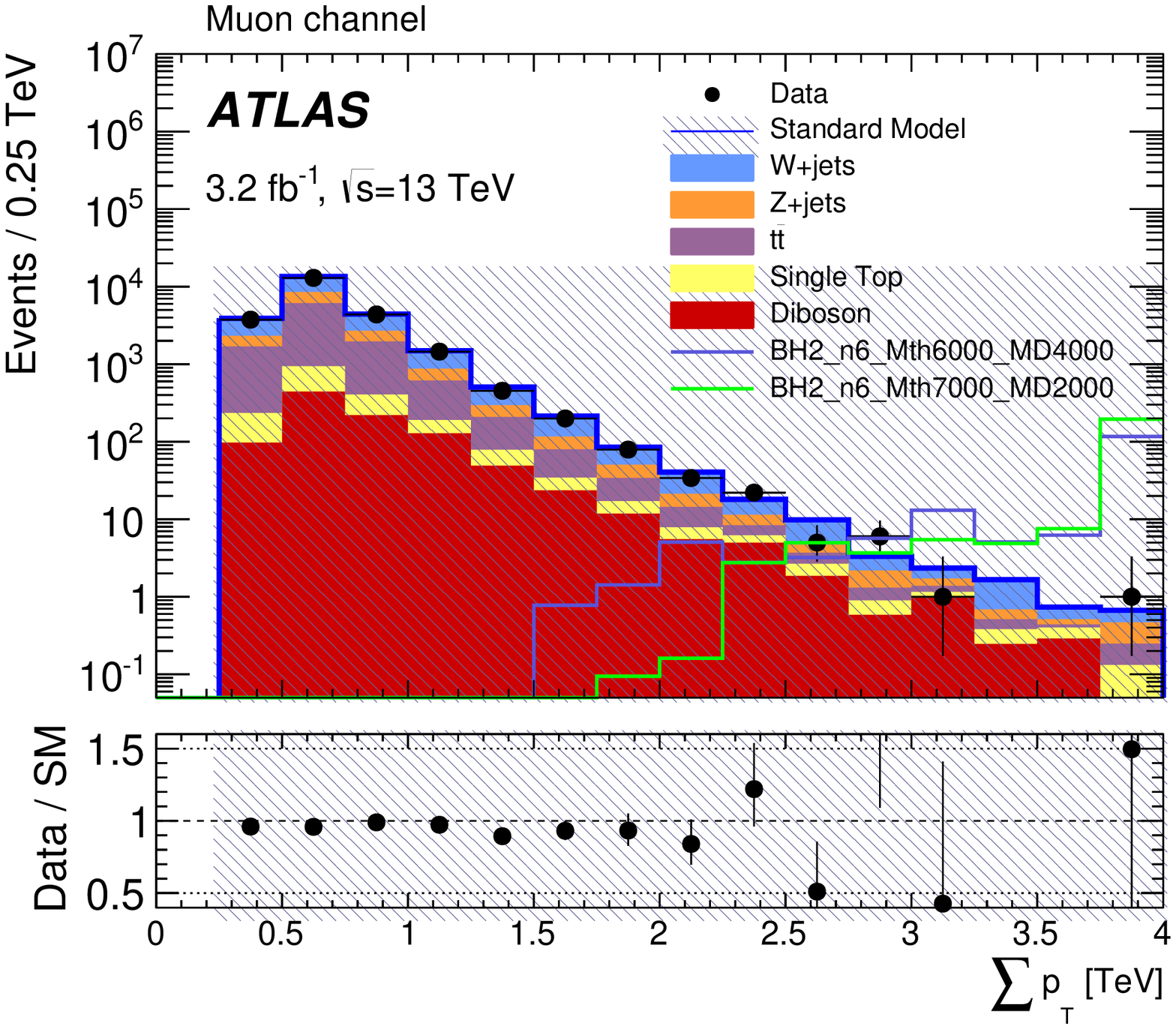}
      \label{fig:finalsumpt:muChan}
    }
    \caption{The  $\sumpt$ distributions in \subref{fig:finalsumpt:elChan} the electron channel and \subref{fig:finalsumpt:muChan} the muon channel.
      The selection is that of the signal regions except for the final requirement on \sumpt.
      \figexplanationwithsignal\ \figdisclaimer\
    }
    \label{fig:finalsumpt}
  \end{center}
\end{figure*}

Figure \ref{fig:finalsumpt} shows the data and background predictions for \sumpt in the electron and muon channels following the background likelihood fit, with two signal models overlaid.
This figure uses the SR selection except for the final requirement on \sumpt.
The data are in good agreement with the background prediction across the range of \sumpt which can be tested with the present data, with the size and pattern of deviations between data and background prediction being consistent with statistical fluctuations and the size of the systematic uncertainties.
Table~\ref{tab:finalResultsShort} presents the data and background predictions in the signal regions.
The number of events observed in \SRTwo\ is higher than the background estimate in the electron channel with a $p$-value of 1\% when tested against the background-only hypothesis.
The excess is not sufficiently significant to be considered as evidence of any new physics effect.
The final results are therefore derived from the combination of the two channels.
The observed numbers of events in \SROne\ and \SRTwo\ are
\observedSrLowwCo\ and
\observedSrHighCo\ respectively
for the combination of the electron and muon channels, to be compared with fitted background predictions of
\expectedSrLowwCo\ and \expectedSrHighCo.

Model-independent cross-section upper limits on any potential new physics contribution are obtained from fits to all control regions and to signal regions combining the electron and muon channels, with potential signal contributions included via a freely-floating parameter in those signal regions.
Model-independent upper limits of
\modelIndependentBoundNinetyFiveLowwCo~\modelIndependentBoundNinetyFiveUnits\ (\modelIndependentBoundNinetyFiveHighCo~\modelIndependentBoundNinetyFiveUnits)
at the 95\% confidence level (CL) are  set on the maximum observable cross-section (defined as cross-section $\times$ acceptance $\times$ efficiency) allowed for any form of new physics in the \SROne\ (\SRTwo) region which produces a lepton in conjunction with at least two other objects, each with $p_\mathrm{T}>100$~GeV.

\begin{figure*}[tbp]
\begin{center}
  \includegraphics[width=0.98\textwidth]{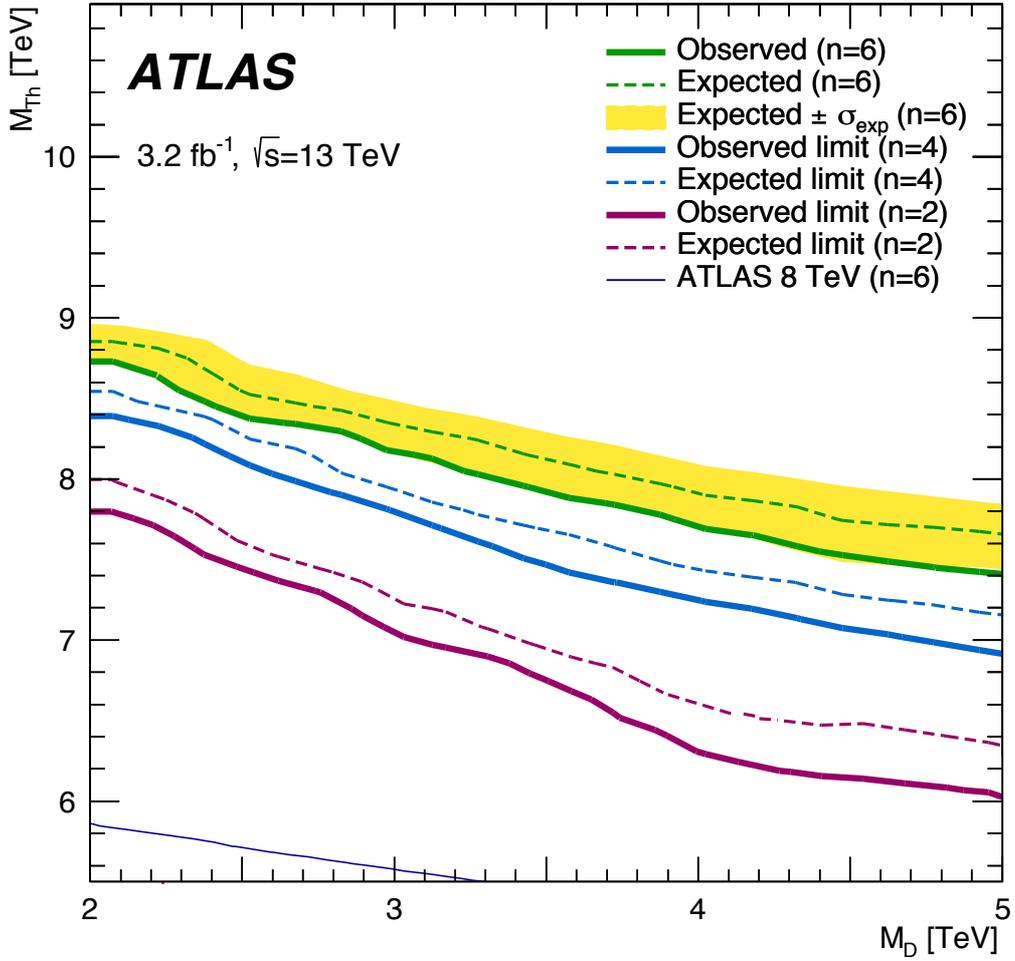}
\caption{Exclusion contours in the \MTH, \MD plane for models of rotating black holes with two, four and six extra dimensions simulated with \charybdis . The solid (dashed) lines show the observed (expected) 95\% CL lower limits, with the yellow (shaded) region illustrating the $\pm 1 \sigma$ variation of the expected limit for six extra dimensions.
The line at the extreme lower left shows the limit set by the analysis at 8~TeV~\cite{Aad:2014gka} for six extra dimensions.
Masses below the corresponding lines are excluded.
\label{fig:BHLimits}}
\end{center}
\end{figure*}

Fits including predicted signal yields in all control and signal regions simultaneously are used to extract exclusion limits for specific black-hole signal models.
Since the signal regions overlap in \sumpt, these exclusion fits are performed for $\sumpt > 3$~TeV, combining the electron and muon data.
Confidence levels are evaluated using the $\mathrm{CL_s}$ procedure~\cite{Read:2002hq}.
The results are shown in Figure \ref{fig:BHLimits}, together with the corresponding limit from the Run~1 analysis at $\sqrt{s} = 8~\text{TeV}$~\cite{Aad:2014gka}. The impact on the \MTH limit for $n=6$ due to the PDF-induced uncertainties in the signal cross-section varies from $\pm 200$~GeV to $\pm 100$~GeV as \MD\ varies from 2~TeV to 5~TeV.
The limit on \MTH\ is more stringent than that from the Run~1 search by almost 3~TeV at $\MD=2$~TeV and by more than 2~TeV at $\MD=4$~TeV.
For a model of rotating black holes with two extra dimensions, the 95\% CL lower limit on the threshold mass \MTH is set at 7.8~TeV for $\MD=2~\text{TeV}$.
For a model with six extra dimensions, the limit is set at 7.4~TeV for $\MD=5~\text{TeV}$.

\FloatBarrier

\begin{table}[tbp]
\begin{center}

\setlength{\tabcolsep}{0.0pc}
\sisetup{table-align-uncertainty=true}
\newcommand{\withoutuncertSRTwoMu}[1]{\multicolumn{1}{S[table-format=1.2]@{\hskip 4mm}}{#1}&\multicolumn{1}{S[table-format=1.2]}{}}
\newcommand{\asymmetricuncertSRTwoMu}[3]{\multicolumn{1}{S[table-format=1.2]@{\hskip 4mm}}{#1}&\multicolumn{1}{l}{$_{#2}^{#3}$}}
\newcommand{\withoutuncertSRTwoEl}[1]{\multicolumn{1}{S[table-format=1.2]@{\hskip 4mm}}{#1}&\multicolumn{1}{S[table-format=1.2]}{}}
\newcommand{\asymmetricuncertSRTwoEl}[3]{\multicolumn{1}{S[table-format=1.2]@{\hskip 4mm}}{#1}&\multicolumn{1}{l}{$_{#2}^{#3}$}}
\newcommand{\withoutuncertWjets}[1]{\multicolumn{1}{S[table-format=2.1]@{\hskip 4mm}}{#1}&\multicolumn{1}{S[table-format=1.1]}{}}
\newcommand{\asymmetricuncertWjets}[3]{\multicolumn{1}{S[table-format=2.1]@{\hskip 4mm}}{#1}&\multicolumn{1}{l}{$_{#2}^{#3}$}}
\newcommand{\withoutuncertSROneMu}[1]{\multicolumn{1}{S[table-format=2.1]@{\hskip 4mm}}{#1}&\multicolumn{1}{S[table-format=1.1]}{}}
\newcommand{\asymmetricuncertSROneMu}[3]{\multicolumn{1}{S[table-format=2.1]@{\hskip 4mm}}{#1}&\multicolumn{1}{l}{$_{#2}^{#3}$}}
\newcommand{\withoutuncertSROneEl}[1]{\multicolumn{1}{S[table-format=3.1]@{\hskip 4mm}}{#1}&\multicolumn{1}{S[table-format=1.1]}{}}
\newcommand{\asymmetricuncertSROneEl}[3]{\multicolumn{1}{S[table-format=3.1]@{\hskip 4mm}}{#1}&\multicolumn{1}{l}{$_{#2}^{#3}$}}
\newcommand{\withoutuncertVR}[1]{\multicolumn{1}{S[table-format=1.2]@{\hskip 4mm}}{#1}&\multicolumn{1}{S[table-format=1.2]}{}}
\newcommand{\asymmetricuncertVR}[3]{\multicolumn{1}{S[table-format=1.2]@{\hskip 4mm}}{#1}&\multicolumn{1}{l}{$_{#2}^{#3}$}}
\newcommand{\withoutuncertttbar}[1]{\multicolumn{1}{S[table-format=3.1]@{\hskip 4mm}}{#1}&\multicolumn{1}{S[table-format=2.1]}{}}
\newcommand{\asymmetricuncertttbar}[3]{\multicolumn{1}{S[table-format=3.1]@{\hskip 4mm}}{#1}&\multicolumn{1}{l}{$_{#2}^{#3}$}}
\newcommand{\withoutuncertZjets}[1]{\multicolumn{1}{S[table-format=1.2]@{\hskip 4mm}}{#1}&\multicolumn{1}{S[table-format=1.2]}{}}
\newcommand{\asymmetricuncertZjets}[3]{\multicolumn{1}{S[table-format=1.2]@{\hskip 4mm}}{#1}&\multicolumn{1}{l}{$_{#2}^{#3}$}}
{\small
\begin{tabular*}{\textwidth}{l @{\hskip 5mm}S[table-format=3.1]@{\hskip 2mm\(\pm\)\hskip 0mm}S[table-format=1.1]@{\hskip 11mm}S[table-format=2.1]@{\hskip 2mm\(\pm\)\hskip 0mm}S[table-format=1.1]@{\hskip 11mm}S[table-format=1.2]@{\hskip 2mm\(\pm\)\hskip 0mm}S[table-format=1.2]@{\hskip 11mm}S[table-format=1.2]@{\hskip 1mm\(\pm\)\hskip 1mm}S[table-format=1.2]@{\hskip 11mm}}
\noalign{\smallskip}\hline\noalign{\smallskip}
{\bf  }           &  \multicolumn{2}{c@{\hskip 4mm}}{\SROne\ (electron)}            &  \multicolumn{2}{c@{\hskip 4mm}}{\SROne\ (muon)}            &  \multicolumn{2}{c@{\hskip 4mm}}{\SRTwo\ (electron)}            &  \multicolumn{2}{c@{\hskip 4mm}}{\SRTwo\ (muon)}              \\[-0.05cm]
\noalign{\smallskip}\hline\noalign{\smallskip}
Observed events          & \withoutuncertSROneEl{123}              & \withoutuncertSROneMu{69}              & \withoutuncertSRTwoEl{11}              & \withoutuncertSRTwoMu{2}                    \\
\noalign{\smallskip}\hline\noalign{\smallskip}
Expected bkg events         & 104 & 9          & 78 & 6          & 4.6 & 0.8          & 5.3 & 1.2              \\
\noalign{\smallskip}\hline\noalign{\smallskip}
        Expected \ttbar\ events         & 13.8 & 3.1          & 11.4 & 2.5          & 0.65 & 0.18          & 0.55 & 0.15              \\
        Expected \Wjets\ events         & 32.0 & 3.5          & 33.9 & 3.2          & 1.76 & 0.31          & 2.0 & 0.4              \\
        Expected \Zjets\ events         & 16.6 & 1.5          & 12.6 & 1.4          & 1.09 & 0.18          & 0.77 & 0.24              \\
        Exp. single-top-quark events         & 6.1 & 0.9          & 5.2 & 0.7          & 0.59 & 0.18          & 0.54 & 0.14              \\
        Expected diboson events         & 11.4 & 1.4          & 14.5 & 1.5          & 0.22 & 0.18          & 1.5 & 0.5              \\
        Expected multijet events         & 24 & 7          & 0.0 & 0.0          & 0.32 & 0.24          & 0.0 & 0.0              \\
 \noalign{\smallskip}\hline\noalign{\smallskip}
\end{tabular*}
}

\caption{ Background fit results for regions \SROne\ ($\sumpt>2$~TeV) and \SRTwo\ ($\sumpt>3$~TeV) for the electron and muons channels.
  \tableerrorsstatement\
}
\label{tab:finalResultsShort}
\end{center}
\end{table}

\FloatBarrier

\section{Conclusion}
\label{sec:conclusion}

A search {has been} performed for signatures of TeV-scale gravity in high-mass final states including at least one lepton in conjunction with at least two other leptons or hadronic jets each with $p_\mathrm{T}>100$~GeV, using \intlumi\ of proton--proton collisions recorded by the ATLAS detector at the LHC at a centre-of-mass energy of 13~TeV. No significant deviation from the background predictions is observed.
Upper limits are therefore set on the possible contribution of new physics processes in this class of final states at
\modelIndependentBoundNinetyFiveLowwCo~\modelIndependentBoundNinetyFiveUnits\ (\modelIndependentBoundNinetyFiveHighCo~\modelIndependentBoundNinetyFiveUnits)
at 95\% CL for $\sumpt>2$~TeV (3~TeV).
Constraints are placed on production of microscopic black holes in models with two to six extra space dimensions which substantially extend the excluded range of model parameters.
The results of this analysis could potentially be used to constrain other models predicting new phenomena at the TeV scale involving decays to leptons and jets.

\section*{Acknowledgements}


We thank CERN for the very successful operation of the LHC, as well as the
support staff from our institutions without whom ATLAS could not be
operated efficiently.

We acknowledge the support of ANPCyT, Argentina; YerPhI, Armenia; ARC, Australia; BMWFW and FWF, Austria; ANAS, Azerbaijan; SSTC, Belarus; CNPq and FAPESP, Brazil; NSERC, NRC and CFI, Canada; CERN; CONICYT, Chile; CAS, MOST and NSFC, China; COLCIENCIAS, Colombia; MSMT CR, MPO CR and VSC CR, Czech Republic; DNRF and DNSRC, Denmark; IN2P3-CNRS, CEA-DSM/IRFU, France; GNSF, Georgia; BMBF, HGF, and MPG, Germany; GSRT, Greece; RGC, Hong Kong SAR, China; ISF, I-CORE and Benoziyo Center, Israel; INFN, Italy; MEXT and JSPS, Japan; CNRST, Morocco; FOM and NWO, Netherlands; RCN, Norway; MNiSW and NCN, Poland; FCT, Portugal; MNE/IFA, Romania; MES of Russia and NRC KI, Russian Federation; JINR; MESTD, Serbia; MSSR, Slovakia; ARRS and MIZ\v{S}, Slovenia; DST/NRF, South Africa; MINECO, Spain; SRC and Wallenberg Foundation, Sweden; SERI, SNSF and Cantons of Bern and Geneva, Switzerland; MOST, Taiwan; TAEK, Turkey; STFC, United Kingdom; DOE and NSF, United States of America. In addition, individual groups and members have received support from BCKDF, the Canada Council, CANARIE, CRC, Compute Canada, FQRNT, and the Ontario Innovation Trust, Canada; EPLANET, ERC, FP7, Horizon 2020 and Marie Sk{\l}odowska-Curie Actions, European Union; Investissements d'Avenir Labex and Idex, ANR, R{\'e}gion Auvergne and Fondation Partager le Savoir, France; DFG and AvH Foundation, Germany; Herakleitos, Thales and Aristeia programmes co-financed by EU-ESF and the Greek NSRF; BSF, GIF and Minerva, Israel; BRF, Norway; Generalitat de Catalunya, Generalitat Valenciana, Spain; the Royal Society and Leverhulme Trust, United Kingdom.

The crucial computing support from all WLCG partners is acknowledged
gratefully, in particular from CERN and the ATLAS Tier-1 facilities at
TRIUMF (Canada), NDGF (Denmark, Norway, Sweden), CC-IN2P3 (France),
KIT/GridKA (Germany), INFN-CNAF (Italy), NL-T1 (Netherlands), PIC (Spain),
ASGC (Taiwan), RAL (UK) and BNL (USA) and in the Tier-2 facilities
worldwide.



%

\printbibliography


\newpage 
\begin{flushleft}
{\Large The ATLAS Collaboration}

\bigskip

M.~Aaboud$^\textrm{\scriptsize 135d}$,
G.~Aad$^\textrm{\scriptsize 86}$,
B.~Abbott$^\textrm{\scriptsize 113}$,
J.~Abdallah$^\textrm{\scriptsize 64}$,
O.~Abdinov$^\textrm{\scriptsize 12}$,
B.~Abeloos$^\textrm{\scriptsize 117}$,
R.~Aben$^\textrm{\scriptsize 107}$,
O.S.~AbouZeid$^\textrm{\scriptsize 137}$,
N.L.~Abraham$^\textrm{\scriptsize 149}$,
H.~Abramowicz$^\textrm{\scriptsize 153}$
% ...
\bigskip
\\
% ...
$^{12}$ Institute of Physics, Azerbaijan Academy of Sciences, Baku, Azerbaijan\\
% ...
$^{64}$ University of Iowa, Iowa City IA, United States of America\\
% ...
$^{86}$ CPPM, Aix-Marseille Universit{\'e} and CNRS/IN2P3, Marseille, France\\
% ...
$^{107}$ Nikhef National Institute for Subatomic Physics and University of Amsterdam, Amsterdam, Netherlands\\
% ...
$^{113}$ Homer L. Dodge Department of Physics and Astronomy, University of Oklahoma, Norman OK, United States of America\\
% ...
$^{117}$ LAL, Univ. Paris-Sud, CNRS/IN2P3, Universit{\'e} Paris-Saclay, Orsay, France\\
% ...
$^{135}$ $^{(a)}$ Facult{\'e} des Sciences Ain Chock, R{\'e}seau Universitaire de Physique des Hautes Energies - Universit{\'e} Hassan II, Casablanca; $^{(b)}$ Centre National de l'Energie des Sciences Techniques Nucleaires, Rabat; $^{(c)}$ Facult{\'e} des Sciences Semlalia, Universit{\'e} Cadi Ayyad, LPHEA-Marrakech; $^{(d)}$ Facult{\'e} des Sciences, Universit{\'e} Mohamed Premier and LPTPM, Oujda; $^{(e)}$ Facult{\'e} des sciences, Universit{\'e} Mohammed V, Rabat, Morocco\\
% ...
$^{137}$ Santa Cruz Institute for Particle Physics, University of California Santa Cruz, Santa Cruz CA, United States of America\\
% ...
$^{149}$ Department of Physics and Astronomy, University of Sussex, Brighton, United Kingdom\\
% ...
$^{153}$ Raymond and Beverly Sackler School of Physics and Astronomy, Tel Aviv University, Tel Aviv, Israel\\
% ...
$^{a}$ Also at Department of Physics, King's College London, London, United Kingdom\\
$^{b}$ Also at Institute of Physics, Azerbaijan Academy of Sciences, Baku, Azerbaijan\\
$^{c}$ Also at Novosibirsk State University, Novosibirsk, Russia\\
$^{d}$ Also at TRIUMF, Vancouver BC, Canada\\
$^{e}$ Also at Department of Physics {\&} Astronomy, University of Louisville, Louisville, KY, United States of America\\
% ...
\end{flushleft}


\end{document}